\documentclass[letterpaper,oneside,11pt]{article}

\usepackage[utf8]{inputenc}
\usepackage{amsthm}
\usepackage{amssymb}
\usepackage{amsmath}
\usepackage{geometry}
\usepackage{commath}
\usepackage{graphicx}
\usepackage{subcaption}
\usepackage{bm}
\usepackage{float}
\usepackage{adjustbox}
\usepackage{caption}
\usepackage{setspace}
\usepackage{multicol}
\usepackage{color}
\usepackage{subfiles}
\usepackage{textcomp}
\usepackage{enumerate}
\usepackage{hyperref}
\usepackage{commath}
\usepackage{mathrsfs}
\usepackage{comment}
\usepackage{cite}
\usepackage{listings}
\usepackage{booktabs}
\usepackage[english]{babel}
\usepackage{multirow}
\usepackage{algorithm}
\usepackage[noend]{algpseudocode}
\usepackage{nicematrix, tikz-cd, tikz}
\usetikzlibrary{matrix}
\usetikzlibrary{shapes,decorations,decorations.pathreplacing,calligraphy,arrows,calc,arrows.meta,fit,positioning}
\tikzset{
    -Latex,auto,node distance =1 cm and 1 cm,semithick,
    state/.style ={ellipse, draw, minimum width = 0.7 cm},
    point/.style = {circle, draw, inner sep=0.04cm,fill,node contents={}},
    bidirected/.style={Latex-Latex,dashed},
    el/.style = {inner sep=2pt, align=left, sloped}
}  

\usepackage{setspace}
\usepackage[compact]{titlesec}
\titlespacing{\section}{0pt}{*0.8}{*0.8}
\titlespacing{\subsection}{0pt}{*0.8}{*0.8}
\titlespacing{\subsubsection}{0pt}{*0.8}{*0.8}

\newcommand{\bu}{ {\boldsymbol u} }

\newcommand{\bx}{ {\boldsymbol x} }

\newcommand{\bz}{ {\boldsymbol z} }

\newcommand{\bbeta}   { {\boldsymbol \beta} }

\newcommand{\btheta}  { {\boldsymbol \theta} }

\title{Supervised Learning of Functional Outcomes with Predictors at Different Scales: A Functional Gaussian Process Approach}
\author{
    R. Jacob Andros \\
    Department of Statistics, Texas A\&M University\\
    Rajarshi Guhaniyogi\\ 
    Department of Statistics, Texas A\&M University\\
    Devin Francom \\
    Los Alamos National Laboratories\\
    Donatella Pasqualini\\
    Los Alamos National Laboratories
    }

\begin{document}

\maketitle

\begin{abstract}
\noindent The analysis of complex computer simulations, often involving functional data, presents unique statistical challenges. Conventional regression methods, such as function-on-function regression, typically associate functional outcomes with both scalar and functional predictors on a per-realization basis. However, simulation studies often demand a more nuanced approach to disentangle nonlinear relationships of functional outcome with predictors observed at multiple scales: domain-specific \emph{functional predictors} that are fixed across simulation runs, and realization-specific \emph{global predictors} that vary between runs.
In this article, we develop a novel supervised learning framework tailored to this setting. We propose an additive nonlinear regression model that flexibly captures the influence of both predictor types. The effects of functional predictors are modeled through spatially-varying coefficients governed by a Gaussian process prior. Crucially, to capture the impact of global predictors on the functional outcome, we introduce a functional Gaussian process (fGP) prior. This new prior jointly models the entire collection of unknown, spatially-indexed nonlinear functions that encode the effects of the global predictors over the entire domain, explicitly accounting for their spatial dependence. This integrated architecture enables simultaneous learning from both predictor types, provides a principled strategies to quantify their respective contributions in predicting the functional outcome, and delivers rigorous uncertainty estimates for both model parameters and predictions. The utility and robustness of our approach are demonstrated through multiple synthetic datasets and a real-world application involving outputs from the Sea, Lake, and Overland Surges from Hurricanes (SLOSH) model.
\textit{Keywords}: Computer simulations; Functional data; Functional Gaussian processes; Spatial statistics; Uncertainty quantification.
\end{abstract}

\section{Introduction}
Advances in sensing, computation, and automated acquisition now enable the routine collection of data indexed over time, space, space–time, or depth across domains such as high-consequence computer simulations, biomedicine, and atmospheric science. As a result, modern datasets increasingly comprise functions observed over continuous domains rather than isolated scalar measurements, motivating statistical frameworks that model how scalar predictors drive functional outcomes. Function-on-scalar regression (FOSR), which relates a functional response to scalar predictors, has become a core methodology in functional data analysis \cite{ramsay2005functional, ferraty2006nonparametric, cao2010future, ainsworth2011functional, lin2017adaptive, guan2020some, jiang2020filtering, cai2021efficient}. FOSR has gained prominence in applications ranging from computer experiments, where functional outputs must be emulated and linked to input parameters \cite{sacks1989design, santner2003design, kennedy2001calibrate, bayarri2007computer, conti2010bayesian}, to imaging analyses that connect subject-level characteristics to brain signals \cite{wang2007support, jeon2025interpretable}. 

In FOSR, the functional response is typically expressed as an additive combination of scalar predictors with smooth, domain-dependent coefficient functions capturing their influence. Estimation strategies commonly expand these coefficient functions in bases (e.g., B-splines, Fourier, wavelets), coupled with regularization to enforce smoothness or enable variable selection \cite{ramsay2005functional, reiss2010fast}. Alternative approaches leverage reproducing kernel Hilbert space methods \cite{wang2022functional} or Bayesian stochastic process priors, most notably Gaussian processes, to encode smoothness and yield principled uncertainty quantification \cite{williams2006gaussian, guhaniyogi2023distributed}. Recent developments further extend FOSR to high-dimensional and structured settings, incorporating penalization, mixed-effects formulations, and scalable inference to accommodate complex, large-scale functional datasets \cite{jiang2020filtering, cai2021efficient}.


While related to prior work, our focus departs in a key respect. Rather than modeling a functional response solely as a function of scalar predictors, we consider a multi-scale predictor structure with two distinct classes:
(a) \emph{functional predictors} that share the response’s domain (e.g., space) and remain fixed across realizations; and (b) scalar-valued \emph{global predictors}, that vary across realizations but are constant over the functional domain. This setting is motivated by physics-based hurricane surge simulations, where a storm’s track and hurricane attributes (e.g., intensity, size, forward speed, landfall location) act as global predictors that change from run to run, while geophysical fields such as elevation are functional predictors defined over the spatial domain and held fixed across simulations. The simulator then produces a functional outcome, such as storm surge or flood depth, over the same spatial domain. Our objective is to characterize the nonlinear, spatially varying influence of both predictor types on the functional outcome and to enable predictive inference. By explicitly modeling how global predictors imprint domain-wide effects and how functional predictors modulate local response behavior, we aim to deliver accurate prediction  and calibrated uncertainty across the entire functional domain.

To this end, we propose a flexible additive regression framework designed to deconstruct these multi-scale effects. The influence of the fixed functional predictors is captured through spatially-varying coefficient functions, each assigned a Gaussian process (GP) prior. This established approach allows their effects to smoothly adapt across the domain, capturing local modulations. The core innovation of our work addresses the more complex influence of the global predictors. Their impact on the regression is represented by a family of unknown, spatially-indexed nonlinear functions, one for each spatial location. To jointly model this entire collection and facilitate robust information sharing, we introduce a novel \emph{functional Gaussian process (fGP)} prior.

The fGP is constructed hierarchically, providing both flexibility and structure. First, each unknown, spatially-indexed nonlinear function (capturing a global predictor's effect) is represented via a fixed basis expansion (e.g., using B-splines). Critically, the coefficients of this basis expansion are not treated as simple scalars, following standard practice. Instead, they are modeled as functional coefficients that vary continuously over the spatial domain.
These functional coefficients are then themselves endowed with independent Gaussian process priors. The hierarchical construction allows the model to capture complex, nonlinear global predictor effects that change smoothly across space. By placing GP priors on the functional coefficients, the model can efficiently borrow strength across different locations in the domain, leading to more stable and robust estimates. The resulting unified Bayesian framework provides a principled means to perform inference with full uncertainty quantification for the effects of both functional and global predictors. This enables not only the estimation of their respective contributions but also the generation of predictions for the entire functional outcome, complete with credible intervals that reflect all sources of uncertainty.

The proposed framework marks a significant departure from traditional function-on-scalar regression (FOSR) models through its explicit inclusion of a fixed, function-valued predictor. Furthermore, it is fundamentally distinct from the growing literature on function-on-function (FoF) regression \cite{preda2007regression, scheipl2015functional, kim2018additive, sun2018optimal, jeon2025deep}. FoF models are designed to relate pairs of random functions, where both the response and predictor vary across realizations. In stark contrast, our setting features a single, fixed functional predictor that is constant for all realizations, alongside a set of scalar predictors that change from one realization to the next.

The central methodological contribution of this work is the formulation of a functional Gaussian process (fGP) to model spatially varying effects of global predictors. Although the term ``functional Gaussian process'' has appeared in prior work—for example, \cite{sung2022functional} consider Gaussian processes whose inputs are entire functions—our construction is conceptually distinct. In our framework, the effects of global predictors are first represented via a set of basis functions, and we then model the joint behavior of the corresponding basis coefficient functions by specifying their spatial covariance structure over the domain. This yields an explicit prior on the collection of effect functions that directly encodes their a priori correlation across locations, providing a different mechanism from approaches that define a Gaussian process on a function space of inputs.

The remainder of this article is organized as follows. Section 2 details the proposed model, including its prior specifications and the computational approach for posterior inference. In Section 3, we evaluate the method's performance through extensive simulation studies. Section 4 then demonstrates the framework's utility as an emulator for the SLOSH hurricane model. Finally, Section 5 offers concluding remarks and outlines directions for future research.


\section{Modeling Spatially-Indexed Functional Outcomes with Functional and Global Predictors}
\subsection{Model Development}
\begin{figure}[h!]
\begin{center}
\begin{tikzpicture}

\node[draw, minimum width=3cm, minimum height=2.35cm, align=center, fill=blue!10] (Z) 
{
$\begin{matrix}
Z_{11} & Z_{12} & \cdots & Z_{1q} \\
Z_{21} & Z_{22} & \cdots & Z_{2q} \\
\vdots & \vdots & \ddots & \vdots \\
Z_{s1} & Z_{s2} & \cdots & Z_{sq}
\end{matrix}$
};
\node[above = 0.1cm of Z, align=center] (Ztext) {$Z \ (S \times q)$ \\ (Global Predictors)};

\node[draw, right=2cm of Z, minimum width=3.2cm, minimum height=2.35cm, align=center, fill=gray!10] (Y)
{
$\begin{matrix}
Y_{1}(u_1) & Y_{1}(u_2) & \cdots & Y_{1}(u_n) \\
Y_{2}(u_1) & Y_{2}(u_2) & \cdots & Y_{2}(u_n) \\
\vdots & \vdots & \ddots & \vdots \\
Y_{S}(u_1) & Y_{S}(u_2) & \cdots & Y_{S}(u_n)
\end{matrix}$
};
\node[above = 0.1cm of Y, align=center] (Ytext) {$Y \ (n \times S)$ \\ (Response)};

\node[draw, minimum width=3.5cm, minimum height=3.9cm, align=center, rotate=90, fill=green!10] at(6.5, -4.7) (X)
{
$\begin{matrix}
x_1(u_1) & x_2(u_1) & \cdots & x_p(u_1) \\
& & & \\
& & & \\
x_1(u_2) & x_2(u_2) & \cdots & x_p(u_2) \\
& & & \\
\vdots & \vdots & \ddots & \vdots \\
& & & \\
x_1(u_n) & x_2(u_n) & \cdots & x_p(u_n)
\end{matrix}$
};
\node[align=center] at(6.5,-8.1) (Xtext) {$X \ (n \times p)$ \\ (Functional Predictors)};

\draw[->, thick] (1.9,0.9) -- (3.85,0.9);
\draw[->, thick] (1.9,0.35) -- (3.85,0.35);
\draw[->, thick] (1.9,-0.75) -- (3.85,-0.75);
\draw[->, thick] (4.65,-2.0) -- (4.65,-1.3);
\draw[->, thick] (6.2,-2.0) -- (6.2,-1.3);
\draw[->, thick] (8.4,-2.0) -- (8.4,-1.3);

\end{tikzpicture}
\caption{How the response variable $Y_s(\bu)$ is constructed from the functional and global predictors. We assume that the data are observed at only $n$ distinct spatial locations in the domain, given by $\mathcal{U}=\{\bu_1,...,\bu_n\}$.}
\label{fig:fgp_data_diagram}
\end{center}
\end{figure}
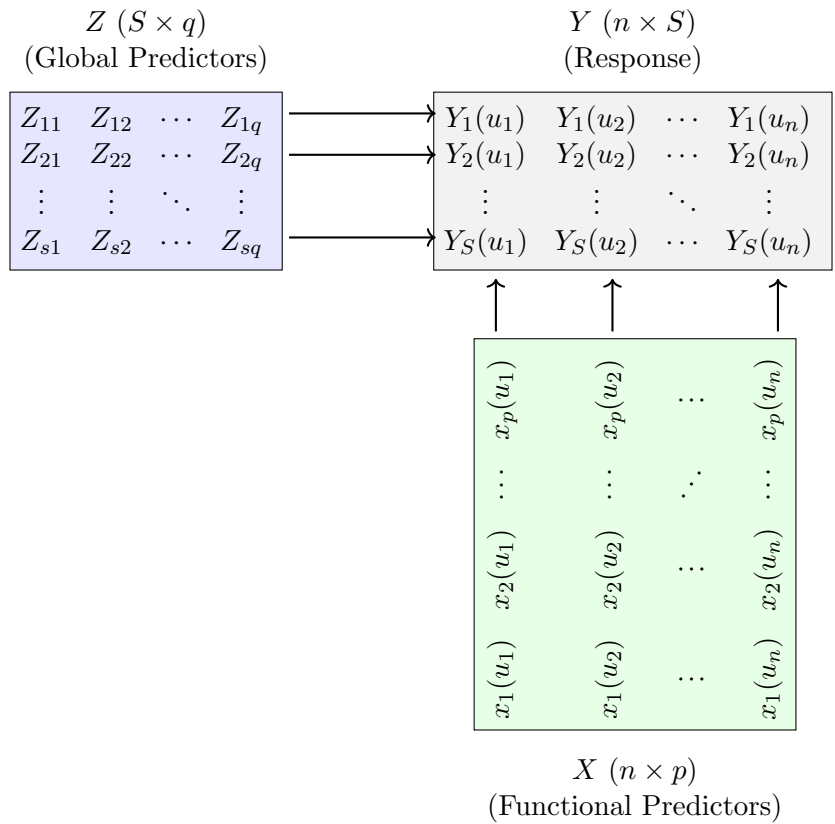

Let $Y_1(\bu), \ldots, Y_S(\bu)$ denote a set of $S$ functional outcomes, or realizations, observed over a compact spatial domain $\mathcal{D} \subseteq \mathbb{R}^2$, with $\bu \in \mathcal{D}$. Each realization $s$ is associated with a vector of $p$ \textbf{global predictors} $\mathbf{z}_s = (z_{s,1}, \ldots, z_{s,p})^T \in \mathbb{R}^p$, which are constant across the domain $\mathcal{D}$. Additionally, a common set of $q$ \textbf{functional predictors} $\mathbf{x}(\mathbf{u}) = (x_1(\mathbf{u}), \ldots, x_q(\mathbf{u}))^T$, which are shared across all realizations, is assumed to influence the outcome. Our objective is to deconstruct how these global and functional predictors jointly shape the functional response. A schematic illustration of the framework is provided in Figure~\ref{fig:fgp_data_diagram}.

To achieve this, we propose an additive, semiparametric regression model that separates the effects of these two predictor types:
\begin{align}\label{eq:additive_local_global}
Y_s(\mathbf{u}) = \underbrace{\mathbf{x}(\mathbf{u})^T\boldsymbol{\beta}(\mathbf{u})}_{\text{Functional Predictor Effect}} + \underbrace{h(\mathbf{z}_s;\bu)}_{\text{Global Predictor Effect}} + \quad \epsilon_s(\mathbf{u}), \quad \mathbf{u} \in \mathcal{D}, \; s = 1, \ldots, S.
\end{align}
Here, $\boldsymbol{\beta}(\mathbf{u})=(\beta_1(\mathbf{u}),\dots,\beta_q(\mathbf{u}))^T \in \mathbb{R}^q$ is a vector of spatially-varying coefficient functions capturing the influence of the functional predictors, while $h(\cdot; \mathbf{u})$ is a nonlinear function modeling the effect of the global predictors at location $\mathbf{u}$. The residual terms are assumed to be i.i.d. Gaussian noise; that is, $\epsilon_s(\mathbf{u}) \sim \mathcal{N}(0, \tau^2)$. While presented for Gaussian responses, this framework readily extends to other data types (e.g., binary, count) by introducing a suitable link function.

This additive structure is a modeling assumption about the underlying process, distinct from ensemble learning strategies (e.g., \cite{freund1997decision}). While both frameworks decompose complex prediction tasks, ensemble methods combine multiple ``weak learners'' to improve predictive performance and reduce overfitting, whereas the additive structure assumes that a high-dimensional regression function can be expressed as a sum of lower-dimensional components. In our case, the additive formulation is a modeling choice rather than a learning strategy. In our formulation, the spatially-varying coefficients $\boldsymbol{\beta}(\mathbf{u})$ offer a flexible yet interpretable way to capture the influence of functional predictors, striking a balance between nonparametric adaptability and parametric clarity. Such models often outperform those with fixed parametric trends (e.g., polynomial regression) by allowing predictor–response relationships to evolve across space \cite{gelfand2003spatial}. We therefore place independent Gaussian process (GP) priors on each coefficient function, $\beta_j(\mathbf{u}) \sim \mathcal{GP}(0, C_\beta(\cdot, \cdot; \boldsymbol{\theta}_{\beta,j}))$ for $j=1,\dots,q$.


\subsection{Functional Gaussian Process Prior on Global Predictor Effects}
The central innovation of our work lies in modeling the effect of the global covariates. Let $\mathcal{H}(\mathcal{D}) = \{h(\cdot;\bu) : \mathbf{u} \in \mathcal{D}\}$ be the uncountable collection of functions that defines this effect across the domain. Our goal is to model the functions in $\mathcal{H}(\mathcal{D})$ jointly, enabling information to be shared across the domain. Specifically, we desire that for locations $\mathbf{u}$ and $\mathbf{u}'$ that are geographically close, the estimation of their corresponding effect functions, $h(\cdot;\bu)$ and $h(\cdot;\bu')$, is strongly related.

To achieve this, we first represent each function $h(\cdot;\bu)$ using a linear combination of $K$ pre-defined basis functions, $B_k(\cdot)$:
\begin{align}\label{basis_expansion}
h(\mathbf{z}_s;\bu) = \sum_{k=1}^K B_k(\mathbf{z}_s)\eta_k(\mathbf{u}), \quad \mathbf{u}\in\mathcal{D}, \quad s=1,\dots,S.
\end{align}
In this formulation, $B_k(\mathbf{z}_s)$ denotes the $k$-th basis function evaluated at the global covariates $\mathbf{z}_s$, and $\eta_k(\bu)$ is the corresponding coefficient, regarded as a function of the spatial location $\bu$. We employ B-splines in our empirical work due to their excellent numerical stability and approximation properties \cite{guhaniyogi2024bayesian}, though other choices like wavelets \cite{vidakovic2009statistical} or radial basis functions \cite{bliznyuk2008bayesian} can also be employed within this framework.

Our framework builds on, yet fundamentally diverges from, the existing literature on basis function expansions. While prior work primarily aims to estimate a single function or a finite set of functions, our goal is to model the entire uncountable collection $\mathcal{H}(\mathcal{D})$ simultaneously. The crucial distinction is that methods like \cite{biller2001bayesian, li2015spatial,bai2019fast} typically treat basis coefficients as scalar random variables, whereas our model treats them as \textbf{spatially-varying random functions}, $\eta_k(\mathbf{u})$.

We place independent GP priors on each coefficient function to flexibly model its spatial variation, $
\eta_k(\cdot) \stackrel{ind.}{\sim} \mathcal{GP}(0, C_k(\cdot, \cdot; \boldsymbol{\theta}_k)), \quad k=1,\dots,K,$
where $\boldsymbol{\theta}_k$ are parameters of the covariance kernel $C_k$. For both $C_k$ and $C_\beta$, we use the Matérn covariance kernel:
\begin{align}\label{eq:matern}
C(\mathbf{u},\mathbf{u}';\boldsymbol{\theta}) = \sigma^2\frac{2^{1-\nu}}{\Gamma(\nu)}\left(\sqrt{2\nu}\frac{\|\mathbf{u}-\mathbf{u}'\|}{\rho}\right)^{\nu} K_{\nu}\left(\sqrt{2\nu}\frac{\|\mathbf{u}-\mathbf{u}'\|}{\rho}\right),
\end{align}
where $\boldsymbol{\theta}=(\sigma^2,\rho,\nu)$ contains the variance, length-scale, and smoothness parameters, respectively. Denote $\btheta_{\beta,j}=(\sigma_{\beta,j}^2,\rho_{\beta,j},\nu_{\beta})$ and $\boldsymbol{\theta}_k=(\sigma_k^2,\rho_k,\nu_k)$.
Following standard practice in spatial statistics \cite{guhaniyogi2022distributed}, we assign Inverse-Gamma priors to variance parameters ($\sigma_k^2, \sigma_{\beta,j}^2$) and Uniform priors to length-scale parameters ($\rho_k, \rho_{\beta,j}$), while pre-specifying the smoothness $\nu_\beta, \nu_1,..,\nu_k$. This specification induces what we refer to as a \textbf{functional Gaussian process (fGP)} prior on the function space $\mathcal{H}(\mathcal{D})$. 

The functional Gaussian process (fGP) prior induces a highly structured and interpretable covariance on the collection of global effect functions, $\mathcal{H}(\mathcal{D})$. Under the fGP prior, the prior covariance between the effect function at location $\mathbf{u}$ for realization $s$ and at location $\mathbf{u}'$ for realization $s'$ is given by
$\mathrm{Cov}(h(\mathbf{z}_s;\bu), h(\mathbf{z}_{s'};\bu'))= C_{\eta,s,s'}(\bu,\bu')= \sum_{k=1}^K B_k(\mathbf{z}_s) C_k(\mathbf{u}, \mathbf{u}'; \boldsymbol{\theta}_k) B_k(\mathbf{z}_{s'}).$
This expression reveals that the prior covariance elegantly decomposes dependence into two distinct components:
\begin{enumerate}
    \item \textbf{Spatial Dependence:} Captured by the kernel $C_k(\mathbf{u}, \mathbf{u}'; \boldsymbol{\theta}_k)$, this component governs how the functions $h(\cdot;\bu)$ and $h(\cdot;\bu')$ are related based on the proximity of their spatial locations. If $\mathbf{u}$ and $\mathbf{u}'$ are close, this term will be large, enforcing that the effect functions themselves are similar.
    
    \item \textbf{Input-Driven Dependence:} Captured by the basis function terms $B_k(\mathbf{z}_s)$ and $B_k(\mathbf{z}_{s'})$, this component modulates the covariance based on the global covariate values for the two different realizations.
\end{enumerate}
To build intuition, consider the special case in which we evaluate the covariance at a \textbf{single fixed location} (\(\mathbf{u}' = \mathbf{u}\)) for two different realizations \(s\) and \(s'\). In this case, the covariance reduces to
$\mathrm{Cov}\big(h(\mathbf{z}_s;\bu), h(\mathbf{z}_{s'};\bu)\big)=C_{\eta,s,s'}(\bu,\bu) 
= \sum_{k=1}^K \sigma_k^2 \, B_k(\mathbf{z}_s)\, B_k(\mathbf{z}_{s'})$,
which shows that, at a fixed location, the covariance between the effects corresponding to two global covariate settings \(\mathbf{z}_s\) and \(\mathbf{z}_{s'}\) is a \textbf{weighted inner product} of their basis expansions. This representation is highly interpretable: when two simulation runs have similar global covariates (\(\mathbf{z}_s \approx \mathbf{z}_{s'}\)), their basis coefficients are also similar, yielding a large positive covariance between their effects. Thus, the model naturally encodes the prior belief that similar inputs lead to similar effects, with the smoothness and structure of this relationship governed by the choice of basis functions.


The model defined by \eqref{eq:additive_local_global}-\eqref{eq:matern} induces a rich, covariate-dependent covariance structure, as formalized below.

\noindent\textbf{Proposition 1 (Induced Covariance Structure).} \textit{The proposed model implies that for any $s,s'\in\{1,\dots,S\}$ and $\mathbf{u},\mathbf{u}'\in\mathcal{D}$, the prior covariance is given by:}
\begin{align}\label{covariance_kernel}
Cov(Y_s(\bu),Y_{s'}(\bu'))&=\sum_{j=1}^q x_j(\bu)C_\beta(\bu,\bu';\btheta_{\beta,j})x_j(\bu')+\sum_{k=1}^K B_k(\bz_s)C_k(\bu,\bu';\btheta_k)B_k(\bz_{s'})\nonumber\\
&+\tau^2 I(s=s',\bu=\bu').
\end{align}
This result shows that the covariance between any two points in the dataset depends explicitly on the functional and global covariates, inducing a non-stationary, covariate-adjusted dependence structure both within and between realizations.

Notably, if we choose a simple linear basis $B_k(\mathbf{z}_s)=\bz_{s}$, our framework reduces to a spatially-varying coefficient model. However, the use of flexible basis functions allows us to capture complex, nonlinear dependence between global covariates and the functional response. Moreover, the smoothness of each functional outcome $Y_s(\cdot)$ is governed by the smoothness of the coefficient functions $\eta_k(\cdot)$ and $\beta_j(\cdot)$. Since a process with a Matérn kernel of smoothness $\nu_k$ is $\lceil\nu_k-1\rceil$ times mean-square differentiable, each functional outcome $Y_s(\cdot)$ will be $\min\{\lceil\nu_\beta-1\rceil,\lceil\nu_1-1\rceil,\dots,\lceil\nu_K-1\rceil\}$ times differentiable. Our simulations explore this by varying $\nu_k$'s and $\nu_{\beta}$ to produce both non-differentiable, once-differentiable and infinitely differentiable outcomes.

Since the primary goal is to develop functional Gaussian process priors for computer experiments with functional and global predictors, we limit our empirical studies to moderate sample sizes. The methodology can be scaled to larger datasets by replacing the Gaussian processes in $\eta_k(\bu)$ and $\beta_j(\bu)$ with more computationally efficient approximations; see \cite{heaton2019case} and references therein. We offer more discussion in Section 5.


\subsection{Posterior Computation}\label{subsec:posterior_comp}

For computational inference, we consider the model at a discrete set of $n$ spatial locations, $\mathcal{U} = \{\mathbf{u}_1, \ldots, \mathbf{u}_n\}$. We stack the functional outcomes into a single $Sn \times 1$ vector $\mathbf{Y} = (Y_1(\mathbf{u}_1), \ldots, Y_1(\mathbf{u}_n),$ $ \ldots, Y_S(\mathbf{u}_1), \ldots, Y_S(\mathbf{u}_n))^T$. A key advantage of the Gaussian process framework is that we can marginalize out the functions $\boldsymbol{\beta}(\mathbf{u})$ and $\{\eta_k(\mathbf{u}): k=1,..,K\}$ to obtain a marginal Gaussian likelihood for the observed data $\mathbf{Y}$.

The joint posterior distribution of all model parameters is proportional to the product of this marginal likelihood and the specified prior distributions, and is given by,
\[
\mathcal{N}(\mathbf{Y} | \mathbf{0}, \boldsymbol{\Sigma}_Y(\btheta_{\beta,1},..,\btheta_{\beta,q},\btheta_1,..,\btheta_K)\times \prod_{j=1}^q p(\sigma_{\beta,j})p(\rho_{\beta,j})\times\prod_{k=1}^Kp(\sigma_k)p(\rho_k),
\]
where the $Sn \times Sn$ covariance matrix $\boldsymbol{\Sigma}_Y$ has an additive structure that directly mirrors the model components:
\begin{align}\label{eq:full_covariance}
\boldsymbol{\Sigma}_Y(\btheta_{\beta,1},..,\btheta_{\beta,q},\btheta_1,..,\btheta_K) = \boldsymbol{\Sigma}_X(\btheta_{\beta,1},..,\btheta_{\beta,q}) + \boldsymbol{\Sigma}_Z(\btheta_1,..,\btheta_K) + \tau^2\mathbf{I}_{Sn}.
\end{align}
Each component in \eqref{eq:full_covariance} corresponds to a source of variation:
\begin{enumerate}
    \item \textbf{Functional Predictor Covariance ($\boldsymbol{\Sigma}_X$):} The covariance induced by the functional predictors is given by $\boldsymbol{\Sigma}_X = \mathbf{J}_S \otimes \mathbf{C}_\beta$, where $\mathbf{J}_S$ is the $S \times S$ matrix of ones. The $n \times n$ matrix $\mathbf{C}_\beta$ encapsulates the shared spatial dependence, with its $(i,i')$-th entry defined as $\sum_{j=1}^q x_j(\mathbf{u}_i)C_{\beta,j}(\mathbf{u}_i, \mathbf{u}_{i'}; \boldsymbol{\theta}_{\beta,j})x_j(\mathbf{u}_{i'})$. The Kronecker product structure reflects that this component is common to all realizations.
    
    \item \textbf{Global Predictor Covariance ($\boldsymbol{\Sigma}_Z$):} The covariance induced by the fGP prior on the global predictor effects is a block matrix where the $(s,s')$-th block is an $n \times n$ submatrix, given by $\sum_{k=1}^K B_k(\mathbf{z}_s)\mathbf{C}_k(\boldsymbol{\theta}_k) B_k(\mathbf{z}_{s'})$. Here, $\mathbf{C}_k(\boldsymbol{\theta}_k)$ being the kernel matrix for $\eta_k(\mathbf{u})$ evaluated at the locations in $\mathcal{U}$.
    
    \item \textbf{Noise Covariance:} The term $\tau^2\mathbf{I}_{Sn}$ represents the independent observation noise.
\end{enumerate}

Posterior inference proceeds via a Markov chain Monte Carlo (MCMC) algorithm. We construct  Metropolis-Hastings steps for sampling the covariance parameters $(\boldsymbol{\theta}_{\beta,1}, \ldots, \boldsymbol{\theta}_{\beta,q},\\ \boldsymbol{\theta}_1, \ldots, \boldsymbol{\theta}_K)$ and the noise variance ($\tau^2$), which do not have closed-form full conditional distributions. This approach efficiently explores the posterior parameter space.

\begin{figure}[h!]
\begin{center}
\begin{tikzpicture}[
    node distance=2cm,
    box/.style={draw, rounded corners, align=center, inner sep=4pt, text width=50mm},
    longbox/.style={draw, rounded corners, align=center, inner sep=4pt},
    param/.style={draw, circle, minimum size=8mm, inner sep=0pt},
    >=latex
]

\node at (5, 5) {Data};
\node at (5.5, 3) {Parameters};
\node at (5.5, 0.5) {Covariance};
\node at (5.9, -2) {Basis Functions};
\node at (6.1, -4) {Final Distribution};

\node[param, fill=blue!10] (Y) at (-6, 5) {$\mathbf{Y}$};
\node[param, fill=blue!10] (X) at (-3.33, 5) {$\mathbf{X}$};
\node[param, fill=blue!10] (U) at (-0.67, 5) {$\mathcal{U}$}; 
\node[param, fill=blue!10] (Z) at (2, 5) {$\mathbf{Z}$};

\node[param, fill=green!10] (sigb) at (-6, 3) {$\sigma_{\beta,j}^2$};
\node[param, fill=green!10] (thetab) at (-4.5, 3) {$\rho_{\beta,j}$};
\node[param, fill=green!10] (sig) at (-0.5, 3) {$\sigma_k^2$};
\node[param, fill=green!10] (theta) at (1, 3) {$\rho_k$};
\node[param, fill=green!10] (tau2) at (3, 3) {$\tau^2$};

\node[box, fill=red!10] (C.beta.j) at (-5,0.5) {
$C_{\beta,j}(\mathbf{u},\mathbf{u}' ) = $
$\sigma^2_{\beta,j} \exp\{-\rho_{\beta,j} \|\mathbf{u}-\mathbf{u}'\|\}$ };

\node[box, fill=red!10] (C.eta.K) at (1,0.5) {
$C_{k}(\mathbf{u}, \mathbf{u}') = $
$\sigma^2_k \exp\{-\rho_k \|\mathbf{u}-\mathbf{u}'\|\}$ };

\node[box, fill=orange!10] (C.beta) at (-5,-2) {
$C_\beta(\mathbf{u}, \mathbf{u}') =
\sum_j x_j(\bu)C_{\beta,j}(\bu,\bu')x_j(\bu')$};

\node[box, fill=orange!10] (C.eta) at (1,-2) {
$C_{\eta,s,s'}(\mathbf{u}, \mathbf{u}') =
\sum_k B_k(\bz_s) C_{k}(\bu,\bu')B_k(\bz_{s'})$
};

\node[longbox, fill=gray!10] (distr) at (-2,-4) {$\mathbf{Y} \sim N \Big( \mathbf{0}, \
\mathbf{J}_S \otimes \mathbf{C}_\beta + \mathbf{C}_\eta + \tau^2 \mathbf{I}_{nS} \Big)$};

\draw[->, thick] (Y) -- (-8.0,5) -- (-8.0,-4) -- (distr);
\draw[->, thick] (tau2) -- (4.2,3) -- (4.2,-4.8) -- (0.1,-4.8) -- (0.1,-4.5);
\draw[->, thick] (Z) -- (3.9,5) -- (3.9,-2.3) -- (3.6,-2.3);
\draw[->, thick] (U) -- (-2,5) -- (-2,0.25) -- (-2.4,0.25);
\draw[->, thick] (U) -- (-2,5) -- (-2,0.25) -- (-1.6,0.25);
\draw[->, thick] (X) -- (-3.33, 1.15);
\draw[->, thick] (sigb) -- (-6, 1.15);
\draw[->, thick] (thetab) -- (-4.5, 1.15);
\draw[->, thick] (sig) -- (-0.5, 1.15);
\draw[->, thick] (theta) -- (1, 1.15);
\draw[->, thick] (-5,-0.15) -- (-5,-1.4);
\draw[->, thick] (1,-0.15) -- (1,-1.4);
\draw[->, thick] (-4,-2.6) -- (-4,-3.5);
\draw[->, thick] (0,-2.6) -- (0,-3.5);

\end{tikzpicture}
\caption{Overview of all model parameters and data used in the proposed model. For illustration, we consider $\nu_{\beta,j}=\nu_k=1/2$, leading to the exponential covariance function for $C_k(\cdot,\cdot;\btheta_{k})$ and $C_{\beta,j}(\cdot,\cdot;\btheta_{\beta,j})$}
\label{fig:fgp_params_diagram}
\end{center}
\end{figure}
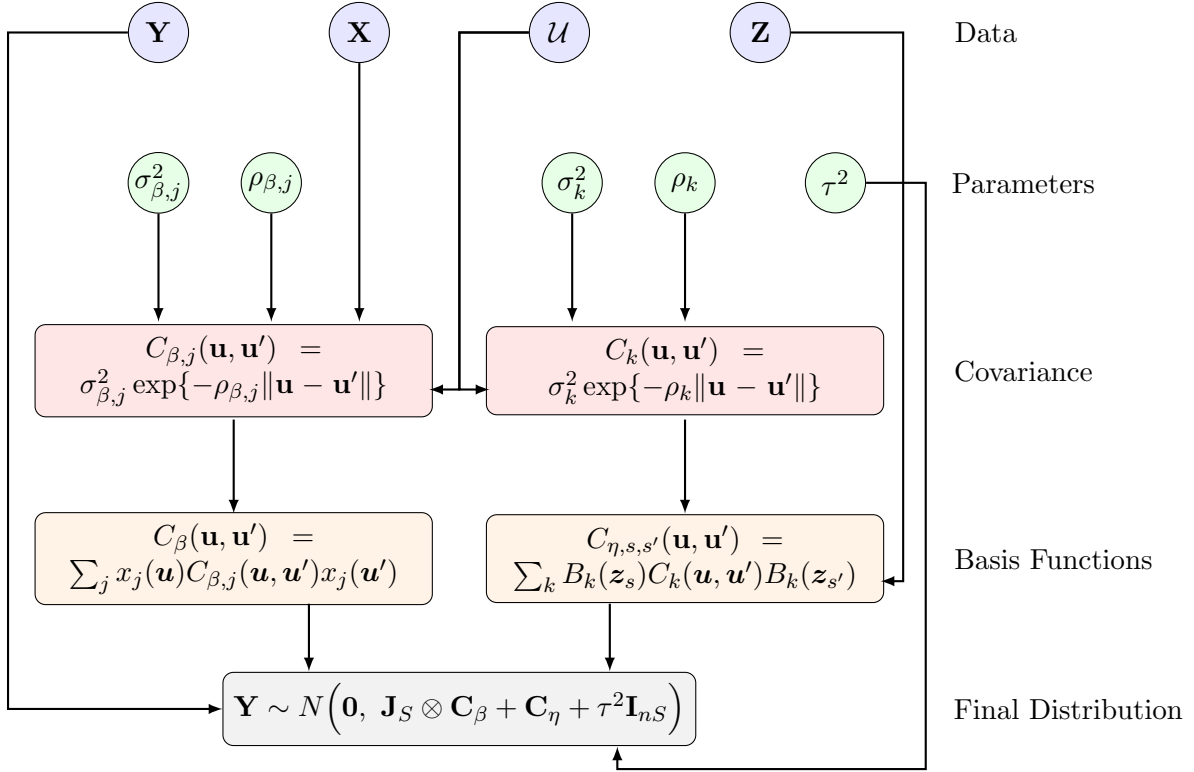




\section{Simulation Studies}
To evaluate the performance of the proposed method, we conduct a series of simulation studies and compare our model with two widely used competitors:  geographically-weighted regression (GWR) and the spatially-varying coefficient (SVC) model. Data are simulated according to the additive functional--global predictor effect formulation in Equation~\eqref{eq:additive_local_global}, and predictive performance is compared across methods.

GWR is employed as a frequentist competitor \cite{fotheringham1998geographically} and is implemented using the R package \texttt{GWmodel} \cite{GWmodel}. For the SVC model, we allow spatially varying
coefficients for both functional and global predictors and fit the model using the R package
\texttt{spBayes} \cite{spBayes}. In all simulation settings, we include two functional
predictors and an intercept, yielding a total of $q = 3$ spatially-varying coefficients. The
local predictor effects are represented using $K = 25$ basis functions, specifically with a tensor
product of cubic B-splines \cite{deboor1978splines} denoted
by $\{B_k(\bz)\}_{k=1}^K$.

All locations are simulated from the square domain of $[0,100]\times [0,100]$.
The $j$th varying coefficient for functional predictors is simulated from a Gaussian process with mean $\mu_j(\bu)$ and covariance kernel $C_{\beta,j}(\bu,\bu';\btheta_{\beta,j})=\sigma_{\beta,j}^2\exp(-\rho_{\beta,j}||\bu-\bu'||)$. 
We explore multiple data-generating scenarios by altering the variance and length-scale parameters
of both functional Gaussian processes and spatially
varying regression coefficients. In particular, we vary $\sigma_1^2,\dots,\sigma_K^2$ and
$\rho_1,\dots,\rho_K$ for the functional Gaussian processes associated with the basis coefficients,
and $\sigma_{\beta,1}^2,\dots,\sigma_{\beta,q}^2$ and $\rho_{\beta,1},\dots,\rho_{\beta,q}$ for the
Gaussian processes governing the spatially-varying coefficient surfaces. The following scenarios are considered:
\begin{itemize}
    \item \textbf{Scenario 1:} Draw $\sigma_{\beta,j}^2$ independently from $[0.5, 1]$ and
    $\rho_{\beta,j}$ from $[0.1, 0.2]$ for $j = 1,\dots,q$. For the fGP processes, draw
    $\sigma_k^2$ from $[5, 10]$ and $\rho_k$ from $[0.1, 0.5]$ for $k = 1,\dots,K$, set the
    nugget variance to $\tau^2 = 0.2$ and the mean functions $\mu_j(\bu)=0$.
    \item \textbf{Scenario 2:} Same as Scenario 1, but with a larger nugget variance
    $\tau^2 = 2$ to reduce signal-to-noise ratio  in the data.
    \item \textbf{Scenario 3:} Same as Scenario 1, except $\rho_{\beta,j}$ are drawn from
    $[0.5, 1]$ to induce smoother coefficient surfaces.
    \item \textbf{Scenario 4:} Same as Scenario 1, but with nonzero mean functions for the spatially varying
    coefficient surfaces:
    $\mu_1(\bu) = u_1 - u_2$,
    $\mu_2(\bu) = u_1 + u_2 - 100$,
    $\mu_3(\bu) = 2u_1 - u_2 - 50$. In this case, the smooth mean structure dominates the local variability, resulting in an overall very smooth spatial surface.
\end{itemize}
The training data uses $S=10$ simulations, each observed over $n=100$ locations. The predictive performance is assessed with another $S_\text{test}=10$ simulations each containing data over $n_\text{test}=25$ locations, under each simulation scenario.

\subsection{Simulation Results}

\begin{figure}[ht]
    \centering
    \begin{subfigure}[b]{0.22\textwidth}
        \centering
        \includegraphics[width=\linewidth]{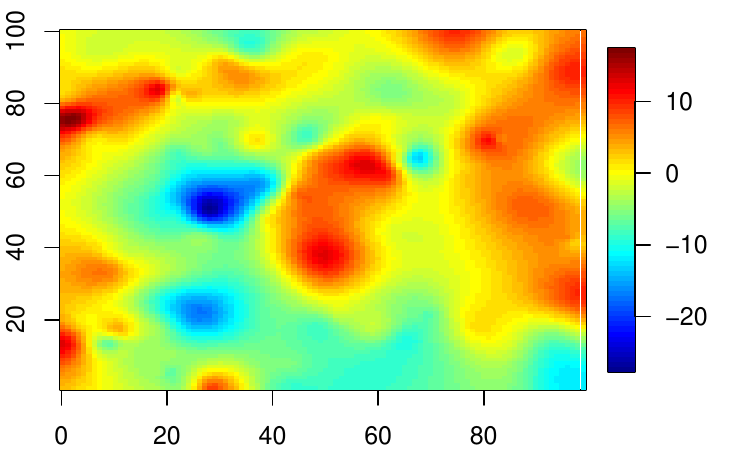}
        \caption*{\footnotesize Scen. 1: true surface}
        \label{fig:subfig1_true}
    \end{subfigure}    
    \begin{subfigure}[b]{0.22\textwidth}
        \centering
        \includegraphics[width=\linewidth]{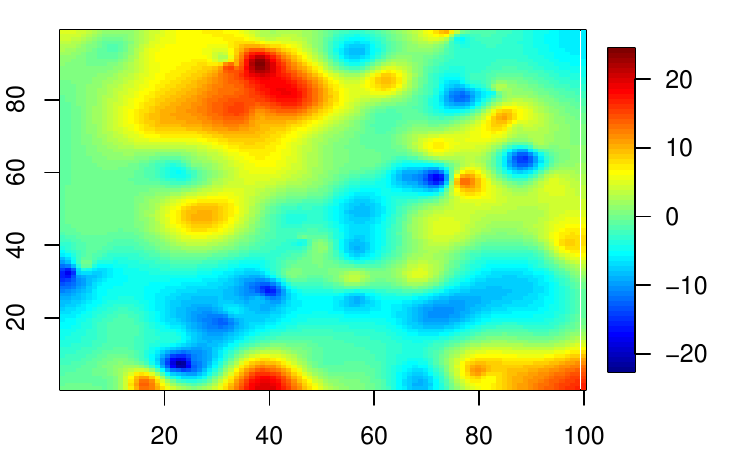}
        \caption*{\footnotesize Scen. 2: true surface}
        \label{fig:subfig2_true}
    \end{subfigure}
    \begin{subfigure}[b]{0.22\textwidth}
        \centering
        \includegraphics[width=\linewidth]{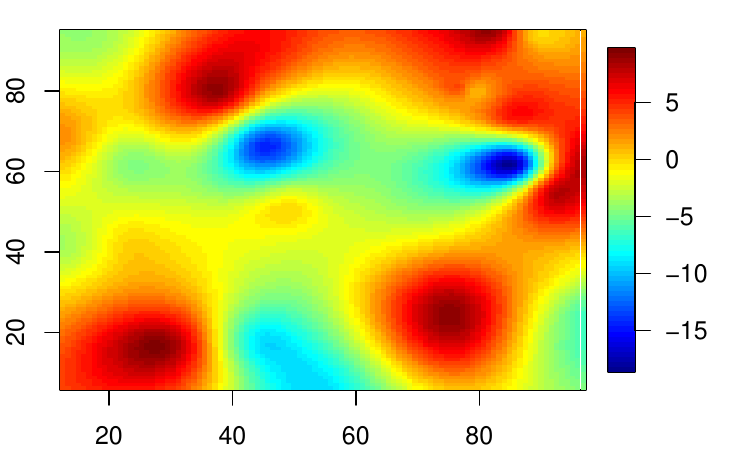}
        \caption*{\footnotesize Scen. 3: true surface}
        \label{fig:subfig3_true}
    \end{subfigure}
    \begin{subfigure}[b]{0.22\textwidth}
        \centering
        \includegraphics[width=\linewidth]{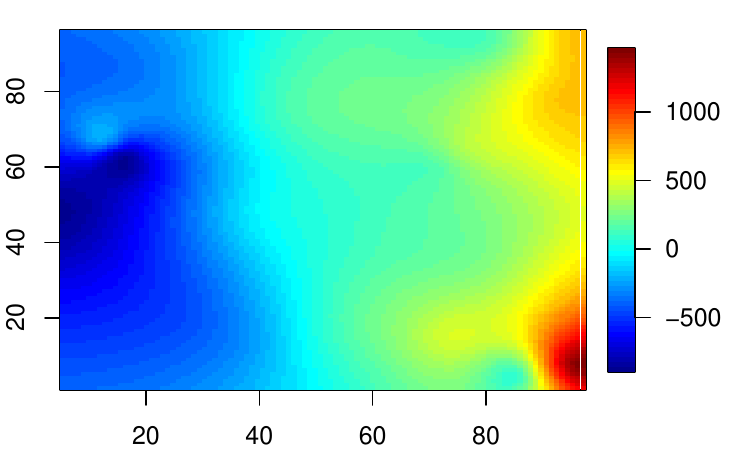}
        \caption*{\footnotesize Scen. 4: true surface}
        \label{fig:subfig4_true}
    \end{subfigure}    
    \begin{subfigure}[b]{0.22\textwidth}
        \centering
        \includegraphics[width=\linewidth]{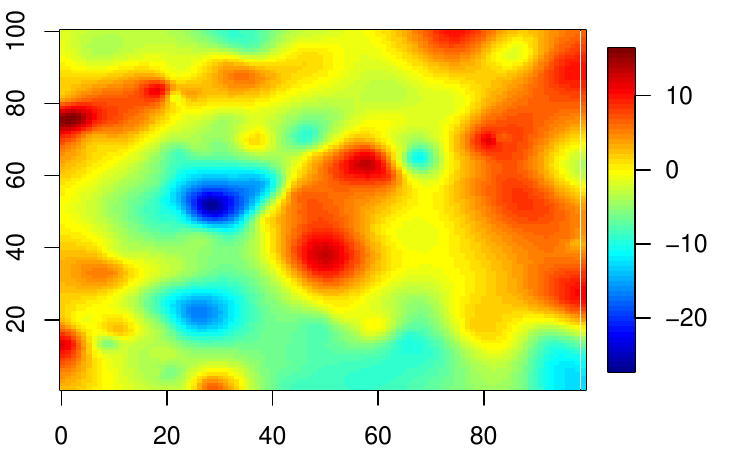}
        \caption*{\scriptsize Scen. 1: predicted surface}
        \label{fig:subfig1_pred}
    \end{subfigure}
    \begin{subfigure}[b]{0.22\textwidth}
        \centering
        \includegraphics[width=\linewidth]{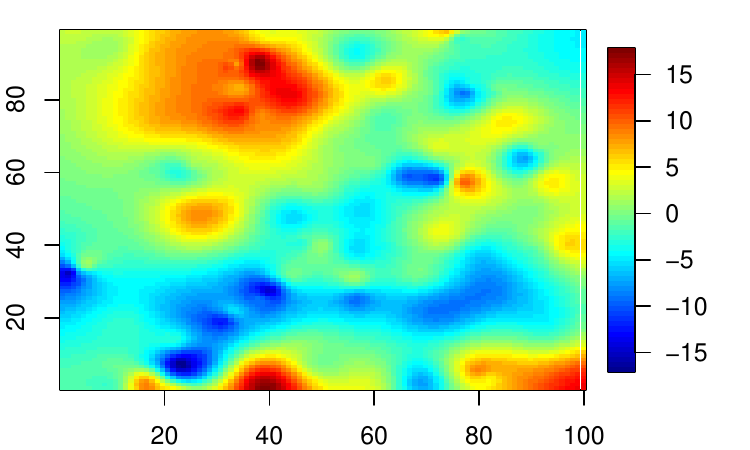}
        \caption*{\scriptsize Scen. 2: predicted surface}
        \label{fig:subfig2_pred}
    \end{subfigure}
    \begin{subfigure}[b]{0.22\textwidth}
        \centering
        \includegraphics[width=\linewidth]{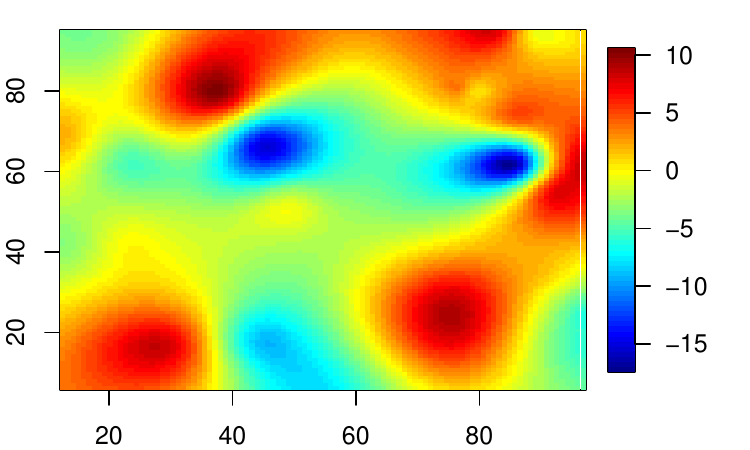}
        \caption*{\scriptsize Scen. 3: predicted surface}
        \label{fig:subfig3_pred}
    \end{subfigure}
    \begin{subfigure}[b]{0.22\textwidth}
        \centering
        \includegraphics[width=\linewidth]{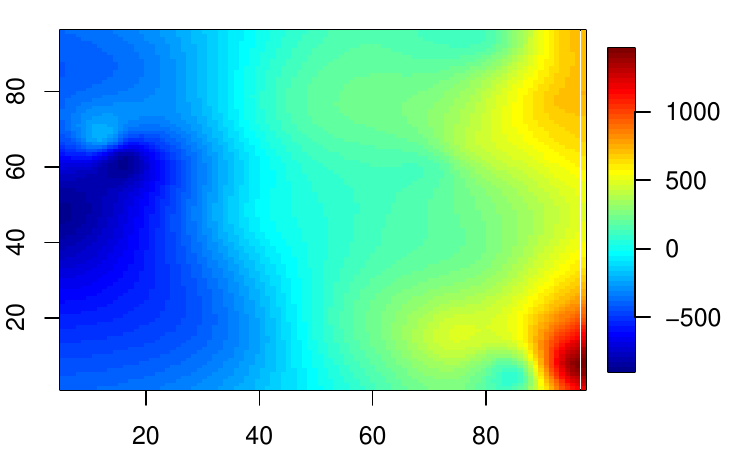}
        \caption*{\scriptsize Scen. 4: predicted surface}
        \label{fig:subfig4_pred}
    \end{subfigure}
    \caption{True and predicted response surfaces produced by the fGP model for a representative out-of-sample simulation from the $S_{test}$ test simulations in Scenarios 1-4.}
    \label{fig:main_figure}
\end{figure}

The surface plots in Figure \ref{fig:main_figure} display both the true and predicted coefficient surfaces for a representative out-of-sample simulation. These plots indicate that the proposed method accurately reproduces the local spatial variability of the underlying surfaces. As the signal-to-noise ratio decreases in Scenario 2, the quality of the predictions degrades mildly, with a few finer-scale features becoming somewhat less distinct. This trend is consistent with the summary results in Table \ref{tab:rmse}, where all competing methods exhibit diminished predictive performance under lower signal-to-noise conditions.

Across all scenarios, fGP substantially outperforms both SVC model and GWR model in terms of root mean squared error (RMSE), as evidenced in Table \ref{tab:rmse}. Regarding uncertainty quantification, fGP achieves empirical coverage for the 95\% predictive intervals that is close to the nominal level, with interval widths that remain reasonably narrow. The main exception occurs in Scenario 2, where the increased noise leads to under-coverage (about 81\%), reflecting the greater difficulty of fully accounting for uncertainty in this case. Consistent with its inaccurate point prediction, SVC shows under-coverage with wider predictive intervals compared to fGP. In contrast, GWR produces overly wide predictive intervals, which, while often nominally well-covered, are so diffuse as to be of limited practical use for inference or decision-making.

\begin{figure}
  \centering
 \begin{subfigure}[b]{0.25\textwidth}
    \includegraphics[width=\textwidth]{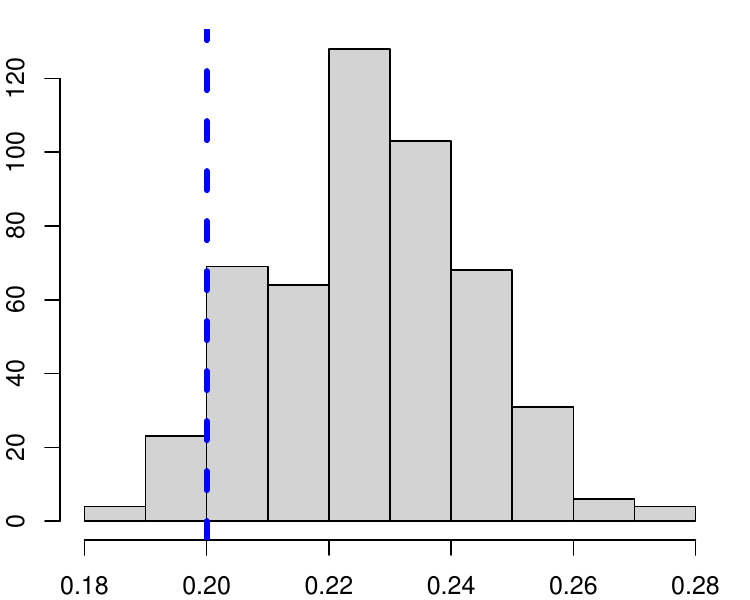}
    \caption*{\footnotesize Scenario 1: posterior of $\tau^2$}
  \end{subfigure}\hfill
  \begin{subfigure}[b]{0.25\textwidth}
    \includegraphics[width=\textwidth]{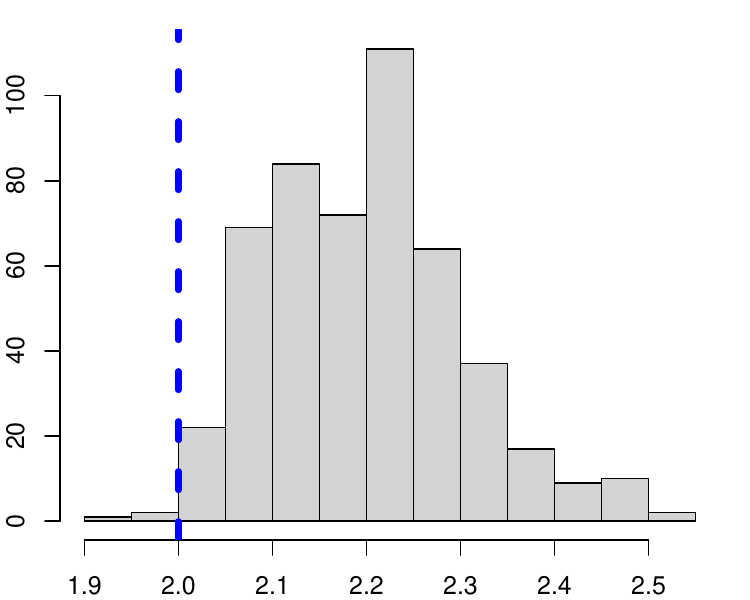}
    \caption*{\footnotesize Scenario 2: posterior of $\tau^2$}
  \end{subfigure}\hfill
  \begin{subfigure}[b]{0.25\textwidth}
    \includegraphics[width=\textwidth]{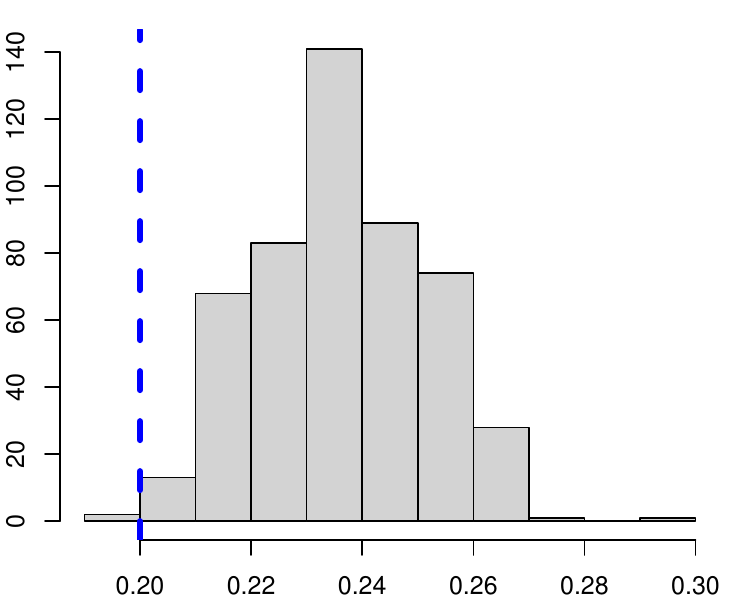}
    \caption*{\footnotesize Scenario 3: posterior of $\tau^2$}
  \end{subfigure}\hfill
  \begin{subfigure}[b]{0.25\textwidth}
    \includegraphics[width=\textwidth]{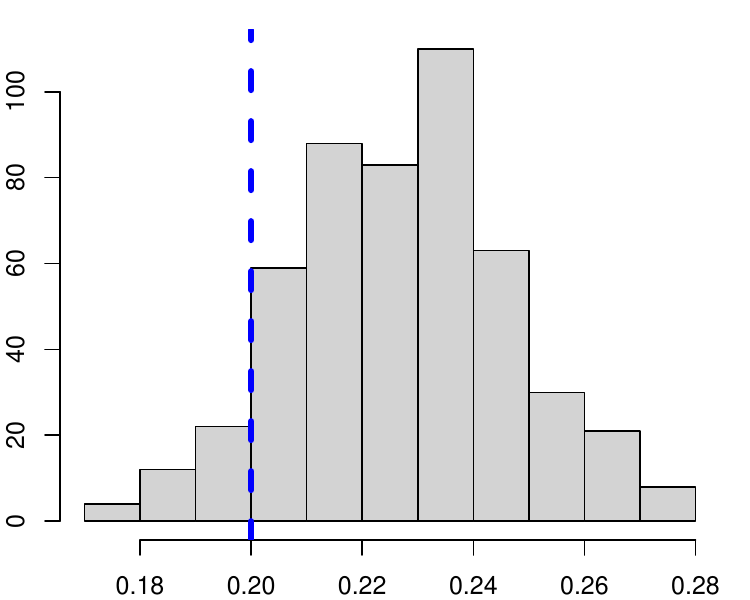}
    \caption*{\footnotesize Scenario 4: posterior of $\tau^2$}
  \end{subfigure}
  \caption{Posterior distributions of $\tau^2$ under Scenarios 1-4. The blue dotted line marks the true value of $\tau^2$. All plots show accurate estimation of the error variance in the four simulation scenarios.}
  \label{fig:tau2_gp}
\end{figure}

Finally, the posterior distribution of the nugget variance $\tau^2$, shown in Figure \ref{fig:tau2_gp}, generally tends to capture the true value (either 0.2 or 2.0, depending on the scenario). This indicates that the model learns the noise level in the data and that the fGP prior in Equation~\ref{eq:additive_local_global}, together with the spatially-varying coefficients, is effectively capturing the contributions of both functional and global predictors. 



\begin{table}[]
    \centering
    \scriptsize
    \begin{tabular}{lrrrrrrrrrrrr}
    \toprule[1.5pt]
        & & \multicolumn{3}{c}{\textbf{fGP}} & & \multicolumn{3}{c}{\textbf{SVC}} & & \multicolumn{3}{c}{\textbf{GWR}} \\
        Scenario & & RMSE & Coverage & Length & & RMSE & Coverage & Length & & RMSE & Coverage & Length \\
        \midrule
        1 (SD $\approx 7.64$) & & 0.74 & 0.95 & 2.88 & & 9.82 & 0.53 & 14.18 & & 8.12 & 0.92 & 28.85 \\
        2 (SD $\approx 8.37$) & & 2.53 & 0.81 & 6.56 & & 9.63 & 0.55 & 15.50 & & 7.64 & 0.92 & 27.33 \\
        3 (SD $\approx 7.10$) & & 0.99 & 0.92 & 3.93 & & 9.10 & 0.47 & 13.71 & & 6.83 & 0.99 & 34.47 \\
        4 (SD $\approx 489$) & & 0.99 & 0.95 & 4.04 & & 59.62 & 0.92 & 229.85 & & 62.22 & 0.92 & 167.30 \\
        \midrule
        5 (SD $\approx 6.66$) & & 2.66 & 0.94 & 11.86 & & 8.05 & 0.43 & 11.26 & & 6.91 & 0.98 & 29.73 \\
        6 (SD $\approx 16.28$) & & 4.76 & 0.73 & 11.93 & & 21.17 & 0.38 & 20.12 & & 17.68 & 0.92 & 62.44 \\
        7 (SD $\approx 8.91$) & & 3.04 & 0.87 & 9.33 & & 10.79 & 0.52 & 14.41 & & 8.60 & 0.91 & 31.27 \\
        8 (SD $\approx 384$) & & 2.20 & 0.95 & 11.37 & & 41.68 & 0.99 & 171.12 & & 43.73 & 0.96 & 177.13 \\
        \bottomrule[1.5pt]
    \end{tabular}
    \caption{Average predictive diagnostics across all out-of-sample simulations. Root mean squared error (RMSE), coverage of 95\% predictive intervals and average length of 95\% predictive intervals are presented for the proposed approach (referred to as fGP) and the competing approaches SVC and GWR. The results show vastly superior point prediction and more precise uncertainty quantification for fGP. The SD values represent the variance of the true responses. RMSE values compared to SD values show that most of the variability of the responses are explained by the model.}
    \label{tab:rmse}
\end{table}

\subsection{Simulation under Model Misspecification}
This section evaluates the performance of the proposed framework under \emph{model misspecification}. In this setting, the data are generated from
\begin{equation}
    Y_s(\bu_i) = \bx_s(\bu_i)^T\bbeta(\bu_i) + g(\bu_i, \bz_s) + \epsilon_s(\bu_i),\:\:\epsilon_s(\bu_i)\stackrel{i.i.d.}{\sim} N(0,\tau^2).
    \label{eq:misspec}
\end{equation}
rather than from the fitted model in \eqref{eq:additive_local_global}. The varying coefficients $\beta_j(\bu)$'s are simulated from the Gaussian processes with mean $\mu_j(\bu)$ and an exponential covariance kernel, as in the previous subsection. In addition, we introduce a function $g(\cdot,\cdot)$ generated from a Gaussian process with mean zero and an exponential covariance function $\text{Cov}(g(\bu_i,\bz_s),g(\bu_{i'},\bz_{s'}))=\zeta^2\exp(-\upsilon\sqrt{||\bu_i-\bu_{i'}||^2+||\bz_s-\bz_{s'}||^2})$.  This misspecified model allows the response variability to arise partly from a latent spatial process $g(\cdot,\cdot)$. The goal is to examine how well our method recovers the overall variability and predictive structure when the assumed model is not exactly correct. We consider the following scenarios by varying 
$\sigma_{\beta,j}^2$, $\rho_{\beta,j}$, $\zeta$, and $\upsilon$:
\begin{itemize}
    \item \textbf{Scenario 5:} $\sigma_{\beta,j}^2$ are independently drawn from $[0.5, 1]$, 
    $\rho_{\beta,j}$ from $[0.1, 0.2]$, and we set $\zeta^2 = 5$, $\upsilon = 0.1$, 
    $\tau^2 = 0.2$, and $\mu_j(\bu) = 0$.
    
    \item \textbf{Scenario 6:} $\sigma_{\beta,j}^2$ are independently drawn from $[3, 5]$, 
    while all other parameters are kept the same as in Scenario 5.
    
    \item \textbf{Scenario 7:} $\rho_{\beta,j}$ are independently drawn from $[0.5, 1]$, 
    with all other parameters the same as in Scenario 5.
    
    \item \textbf{Scenario 8:} All parameters are drawn as in Scenario 5, but with nonzero 
    mean functions for the coefficient surfaces:
    $\mu_1(\bu) = u_1 - u_2$, 
    $\mu_2(\bu) = u_1 + u_2 - 100$, and 
    $\mu_3(\bu) = 2u_1 - u_2 - 50$.
\end{itemize}
These scenarios are designed to evaluate the robustness of the proposed approach
under varying levels of spatial dependence in the functional outcomes and 
different signal-to-noise regimes.

\begin{figure}[ht]
    \centering
    \begin{subfigure}[b]{0.22\textwidth}
        \centering
        \includegraphics[width=\linewidth]{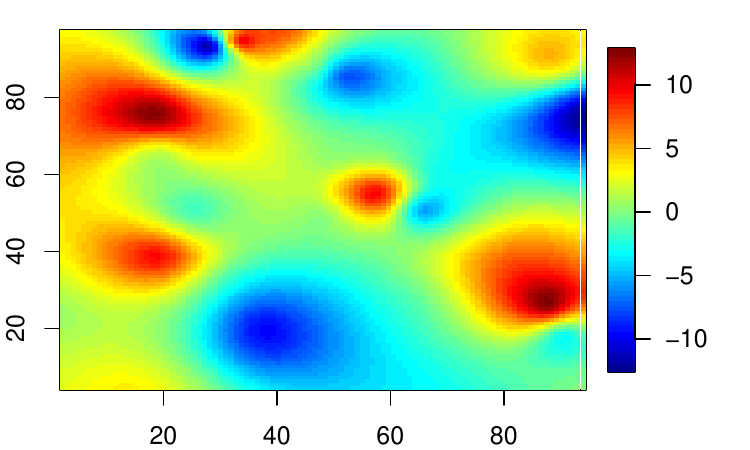}
        \caption*{\footnotesize Scen. 5: true surface}
        \label{fig:subfig5_true}
    \end{subfigure}    
    \begin{subfigure}[b]{0.22\textwidth}
        \centering
        \includegraphics[width=\linewidth]{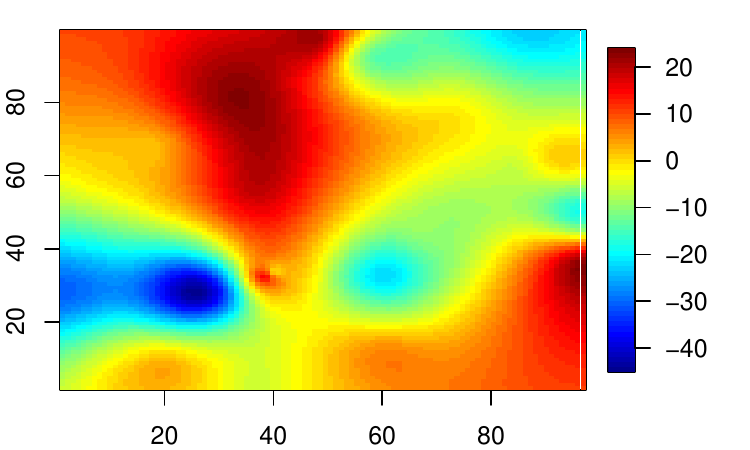}
        \caption*{\footnotesize Scen. 6: true surface}
        \label{fig:subfig6_true}
    \end{subfigure}
    \begin{subfigure}[b]{0.22\textwidth}
        \centering
        \includegraphics[width=\linewidth]{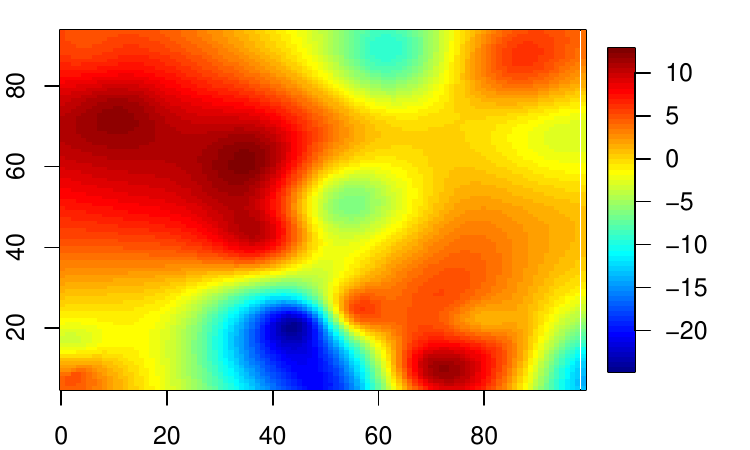}
        \caption*{\footnotesize Scen. 7: true surface}
        \label{fig:subfig7_true}
    \end{subfigure}
    \begin{subfigure}[b]{0.22\textwidth}
        \centering
        \includegraphics[width=\linewidth]{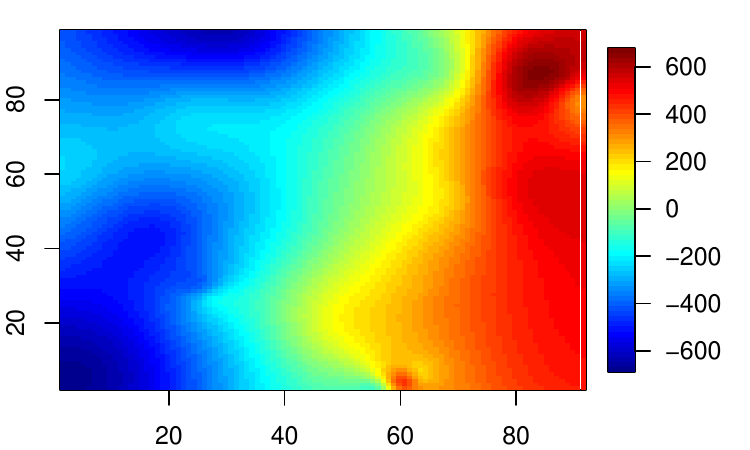}
        \caption*{\footnotesize Scen. 8: true surface}
        \label{fig:subfig8_true}
    \end{subfigure}    
    \begin{subfigure}[b]{0.22\textwidth}
        \centering
        \includegraphics[width=\linewidth]{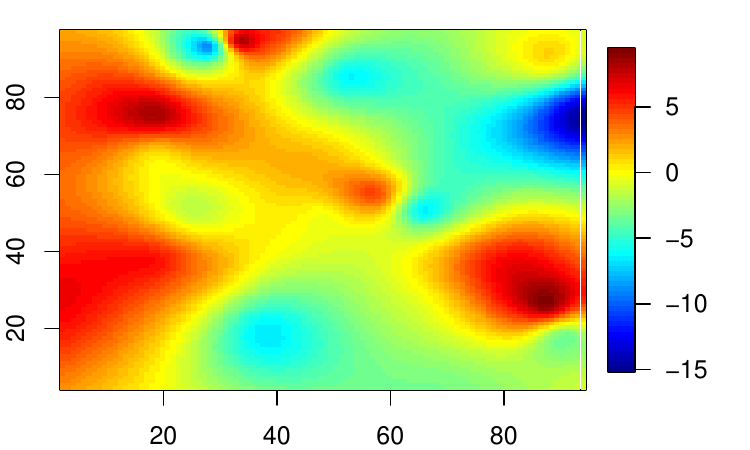}
        \caption*{\scriptsize Scen. 5: predicted surface}
        \label{fig:subfig5_pred}
    \end{subfigure}
    \begin{subfigure}[b]{0.22\textwidth}
        \centering
        \includegraphics[width=\linewidth]{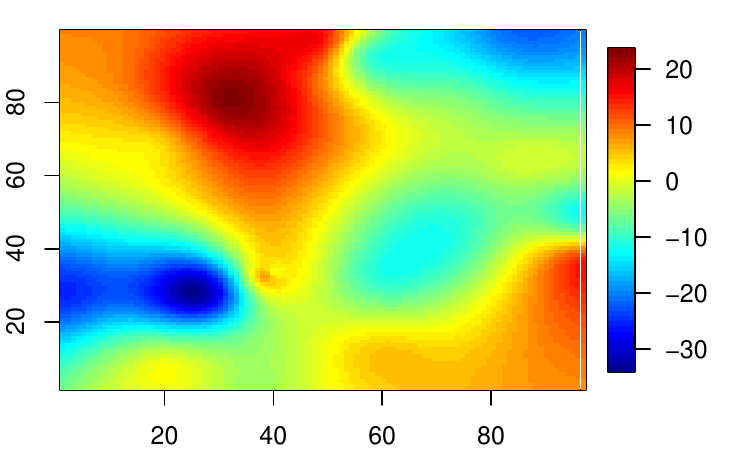}
        \caption*{\scriptsize Scen. 6: predicted surface}
        \label{fig:subfig6_pred}
    \end{subfigure}
    \begin{subfigure}[b]{0.22\textwidth}
        \centering
        \includegraphics[width=\linewidth]{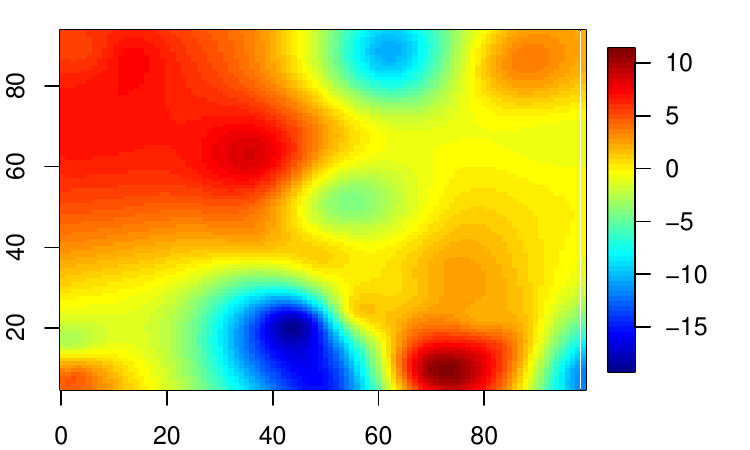}
        \caption*{\scriptsize Scen. 7: predicted surface}
        \label{fig:subfig7_pred}
    \end{subfigure}
    \begin{subfigure}[b]{0.22\textwidth}
        \centering
        \includegraphics[width=\linewidth]{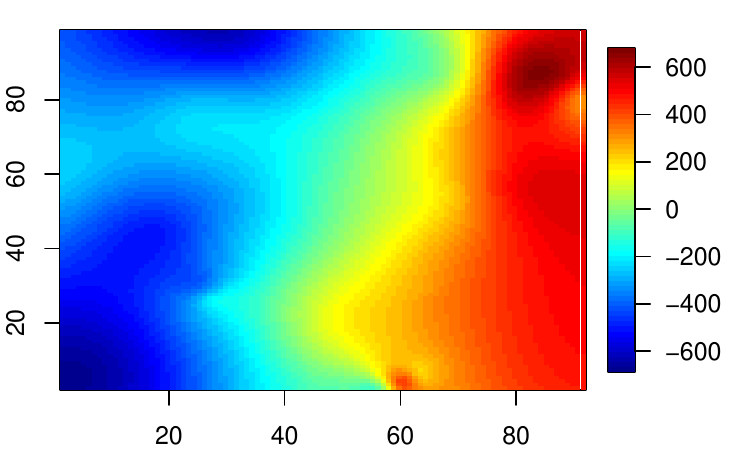}
        \caption*{\scriptsize Scen. 8: predicted surface}
        \label{fig:subfig8_pred}
    \end{subfigure}
    \caption{True and predicted response surfaces produced by the fGP model for a representative out-of-sample simulation from the $S_{test}$ test simulations in Scenarios 5–8, which illustrate cases of model misspecification.}
    \label{fig:main_figure_mis}
\end{figure}

Figure \ref{fig:main_figure_mis} illustrates that, across all scenarios of model misspecification, the response surface estimated by the fGP model for a representative out-of-sample simulation closely aligns with the true underlying surface. This close correspondence demonstrates strong point prediction capabilities of fGP, as further supported by the RMSE values reported in Table \ref{tab:rmse}. Notably, the fGP consistently achieves substantially lower RMSE values compared to alternative methods such as GWR and SVC, highlighting its superior predictive performance even when the model is misspecified.

The standard deviation (SD) of the predicted response in Table~\ref{tab:rmse} indicates that fGP successfully captures and explains a large portion of the variability in the data, which further confirms its robust modeling capacity in challenging settings. Regarding uncertainty quantification, fGP tends to show some degree of under-coverage in the presence of model misspecification under Scenarios 6 and 7. This under-coverage is likely attributable to the use of a finite number of basis functions in the fGP prior when modeling the function $g(\bu_i,\bz_s)$ generated from a full GP, which may restrict the model’s ability to fully account for unexplained uncertainty \cite{stein2014limitations}. In contrast, GWR generally produce overly conservative predictive intervals, characterized by over-coverage and intervals that are excessively wide. SVC consistently shows severe under-coverage except in Scenario 8, characterized by extremely smooth response surface. Overall, the results highlight the effectiveness of fGP in accurately predicting complex response surfaces, even in cases where the true underlying model diverges from the one specified, and underscore the importance of ongoing methodological development for credible uncertainty assessment.
\section{SLOSH Data Analysis}

The Sea, Lake, and Overland Surges from Hurricanes (SLOSH) simulator is a numerical model developed by the National Weather Service to estimate storm-surge--induced floodwater depths associated with tropical cyclones and hurricanes \cite{jelesnianski1992slosh}. Such storm surge models are routinely used to support emergency preparedness, evacuation planning, and real-time response in vulnerable coastal regions. In this study, we focus on the southern New Jersey shoreline, one of the coastal areas for which SLOSH output is available.

For each simulated storm, the SLOSH dataset provides several global predictors that describe large-scale storm characteristics at landfall, including the heading of the storm’s eye, the translation speed of the eye, the latitude at landfall, and the minimum central pressure. To enrich this information with location-specific geographic features, we include local altitude (elevation above mean sea level) as a functional covariate. This functional predictor is defined over the spatial domain of interest and is intended to capture how topography modulates storm surge impacts.

The complete SLOSH output for southern New Jersey is high-dimensional, but large-scale computation is not the main focus of this article. Consequently, we restrict attention to a smaller  subset of spatial locations drawn from the central region of the Cape May peninsula. For model fitting, we use $n=100$ training locations and $S=10$ storms. Predictive performance is then assessed using $S_{test}=10$ additional storms, each observed at $n_{test}=25$ test locations.

We apply our proposed fGP model to this SLOSH subset and compare its performance to SVC and GWR as in the simulation studies. 
Predictive performance on the held-out test locations for all the simulations is summarized in Table~\ref{tab:slosh}. The table reports the root mean squared error (RMSE) of point predictions, along with empirical coverage and average width of nominal 95\% prediction intervals, where available. 

\begin{table}[t]
    \centering
    \begin{tabular}{rrrr}
    \toprule[1.5pt]
         & RMSE & 95\% Int.\ Coverage & 95\% Int.\ Width \\
         \midrule
         fGP & 0.66 & 0.95 & 8.04 \\
         SVC & 1.19 & 0.58 & 2.65 \\
         GWR & 0.94 & 0.66 & 1.95 \\
         \bottomrule[1.5pt]
    \end{tabular}
    \caption{Prediction performance on the SLOSH subset for the proposed model, referred to as fGP, along with competitors SVC and GWR. We present root mean squared error (RMSE), coverage of the 95\% predictive intervals and average length of the 95\% predictive intervals. The results show superior point prediction of fGP compared to SVC and GWR, along with nominal coverage. For reference, the empirical standard deviation of the test responses is 2.18.}
    \label{tab:slosh}
\end{table}

The fGP model achieves the lowest RMSE, substantially improving on both SVC and GWR relative to the test-set variability (standard deviation of 2.18). In terms of uncertainty quantification, fGP produces relatively wide intervals but attains the desired 95\% coverage on average, indicating well-calibrated predictive uncertainty. By contrast, SVC yields considerably narrower intervals but only 58\% empirical coverage, suggesting that its predictive uncertainty is underestimated. 

\begin{figure}[t]
  \centering
  \adjustbox{valign=M}{\includegraphics[height=6cm]{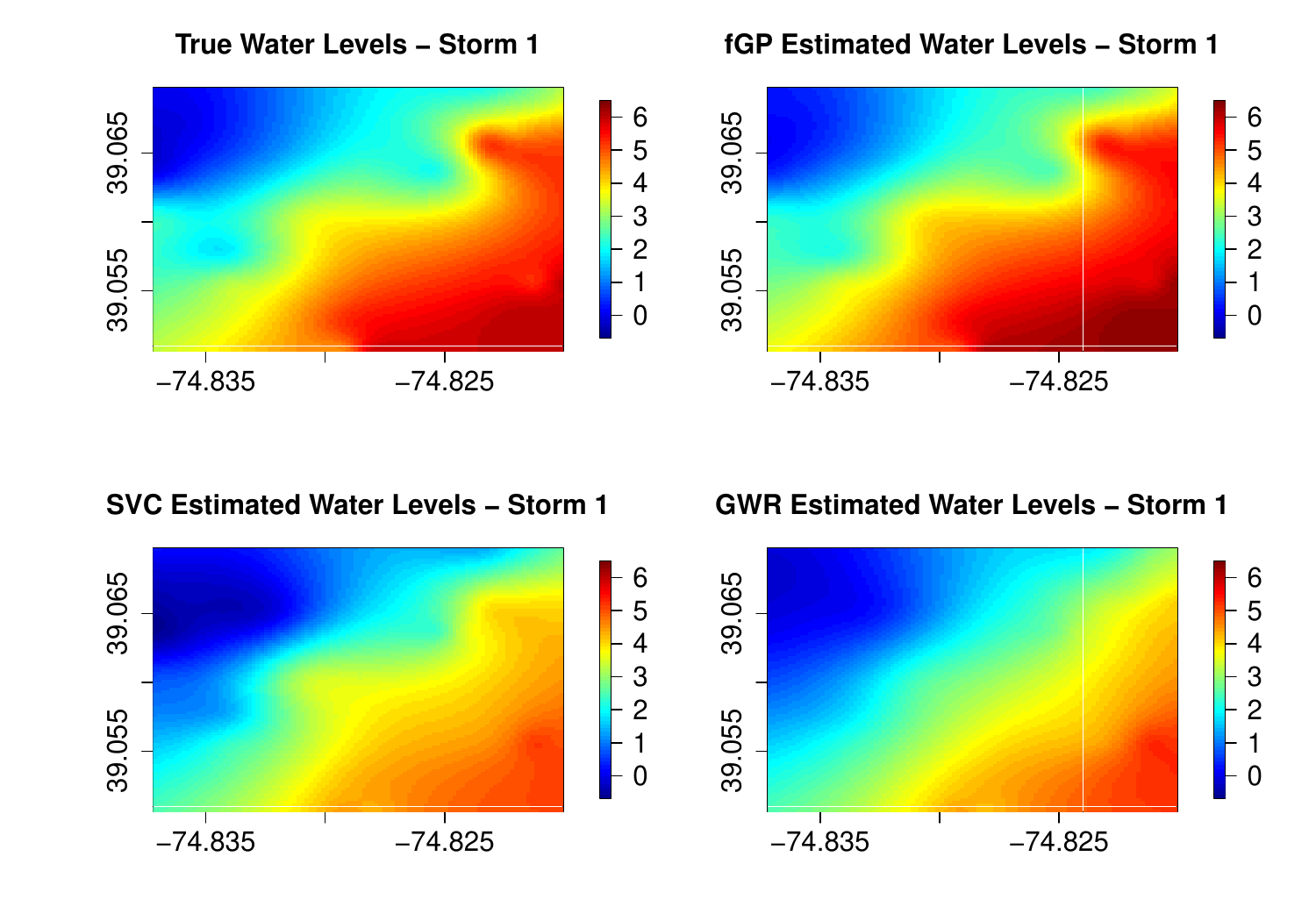}}
  \hspace{1cm}
  \adjustbox{valign=M}{\includegraphics[height=5cm]{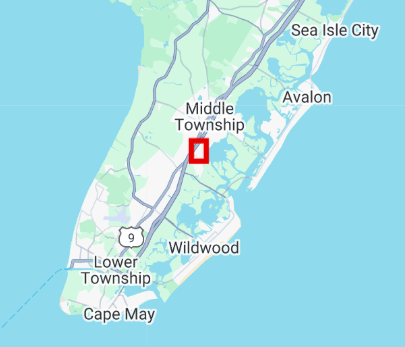}}
  \caption{True water level for the first test storm at all test coordinates, along with predictions from the fGP model and the two competitors (SVC and GWR). The red rectangle in the map (right) indicates the region of New Jersey’s southern peninsula from which the SLOSH subset was drawn.}
  \label{fig:slosh_surface_plots}
\end{figure}

Although we cannot display results for all $S_{\text{test}} = 10$ storms, Figure~\ref{fig:slosh_surface_plots} presents representative surface plots for the first test storm. The left panels compare the true water levels at the test locations to the predicted surfaces from fGP, SVC, and GWR, while the right panel shows the geographic context: the red rectangle on the map of New Jersey highlights the region of the southern peninsula from which the spatial locations were selected.

Across models, all methods capture the relatively low water levels observed in the northwestern (upper-left) portion of the test domain. However, SVC and GWR systematically underestimate the higher storm surge depths in the southeastern (lower-right) region, where water tends to accumulate more strongly. The fGP predictions better track these elevated water levels, leading to improved overall accuracy and more reliable characterization of the spatially varying storm surge response.

\section{Conclusion and Future Work}

In the setting of a functional outcome, the simultaneous presence of functional and global predictors calls for a more flexible modeling framework. Existing approaches such as GWR typically represent the functional outcome as an additive function of scalar covariates. In contrast, our method introduces a multi-scale predictor structure that incorporates both functional and global predictors. As in standard SVC models, we use spatially-varying coefficient functions endowed with Gaussian process priors to capture the effects of the fixed functional predictors. The key novelty lies in how we model the contribution of global predictors: we represent their effects through a collection of unknown nonlinear functions that vary over space, and we place a new functional Gaussian process (fGP) prior on this infinite family of spatially indexed functions.

To assess the practical utility of the fGP, we conducted extensive simulation studies under a variety of conditions, systematically varying spatial variance, noise variance, and smoothness. We also examined performance under deliberate model misspecification. In addition, we analyzed water level outputs from the Sea, Lake, and Overland Surges from Hurricanes (SLOSH) emulator to illustrate the application of the fGP in a realistic scientific problem. Across both the simulations and the SLOSH case study, the fGP framework consistently yielded improved point predictions and more reliable uncertainty quantification compared with competing approaches such as SVC and GWR.

From a computational standpoint, the proposed framework can become demanding when the number of simulations or spatial locations is large, and additional methodological developments will be needed for the fGP to scale efficiently to very large datasets. Because this article is primarily concerned with introducing the fGP prior and demonstrating how to model nonlinear effects of global and functional predictors, we restrict our empirical work to moderate sample sizes and a moderate number of simulations. A promising direction for future research is to employ scalable GP approximations for the spatially-varying coefficients and basis coefficient functions, following \cite{heaton2019case} and references therein.

\bibliographystyle{plain} 
\bibliography{references}

@article{stein2014limitations,
  title={Limitations on low rank approximations for covariance matrices of spatial data},
  author={Stein, Michael L},
  journal={Spatial Statistics},
  volume={8},
  pages={1--19},
  year={2014},
  publisher={Elsevier}
}

@article{reiss2010fast,
  title={Fast function-on-scalar regression with penalized basis expansions},
  author={Reiss, Philip T and Huang, Lei and Mennes, Maarten},
  journal={The international journal of biostatistics},
  volume={6},
  number={1},
  year={2010},
  publisher={De Gruyter}
}

@misc{santner2003design,
  title={The Design and analysis of computer experiments},
  author={Santner, TJ},
  year={2003},
  publisher={Springer}
}

@article{heaton2019case,
  title={A case study competition among methods for analyzing large spatial data},
  author={Heaton, Matthew J and Datta, Abhirup and Finley, Andrew O and Furrer, Reinhard and Guinness, Joseph and Guhaniyogi, Rajarshi and Gerber, Florian and Gramacy, Robert B and Hammerling, Dorit and Katzfuss, Matthias and others},
  journal={Journal of Agricultural, Biological and Environmental Statistics},
  volume={24},
  pages={398--425},
  year={2019},
  publisher={Springer}
}

@book{jelesnianski1992slosh,
  title={SLOSH: Sea, lake, and overland surges from hurricanes},
  author={Jelesnianski, Chester P},
  volume={48},
  year={1992},
  publisher={US Department of Commerce, National Oceanic and Atmospheric Administration~…}
}

@article{wang2022functional,
  title={Functional linear regression with mixed predictors},
  author={Wang, Daren and Zhao, Zifeng and Yu, Yi and Willett, Rebecca},
  journal={Journal of Machine Learning Research},
  volume={23},
  number={266},
  pages={1--94},
  year={2022}
}

@article{bai2019fast,
  title={Fast algorithms and theory for high-dimensional Bayesian varying coefficient models},
  author={Bai, Ray and Boland, Mary R and Chen, Yong},
  journal={arXiv preprint arXiv:1907.06477},
  year={2019}
}

@article{jeon2025deep,
  title={Deep Generative Modeling with Spatial and Network Images: An Explainable AI (XAI) Approach},
  author={Jeon, Yeseul and Guhaniyogi, Rajarshi and Scheffler, Aaron},
  journal={arXiv preprint arXiv:2505.12743},
  year={2025}
}

@article{guhaniyogi2023distributed,
  title={Distributed Bayesian inference in massive spatial data},
  author={Guhaniyogi, Rajarshi and Li, Cheng and Savitsky, Terrance and Srivastava, Sanvesh},
  journal={Statistical science},
  volume={38},
  number={2},
  pages={262--284},
  year={2023},
  publisher={Institute of Mathematical Statistics}
}

@book{williams2006gaussian,
  title={Gaussian processes for machine learning},
  author={Williams, Christopher KI and Rasmussen, Carl Edward},
  volume={2},
  number={3},
  year={2006},
  publisher={MIT press Cambridge, MA}
}

@article{jeon2025interpretable,
  title={Interpretable Deep Neural Network for Modeling Functional Surrogates},
  author={Jeon, Yeseul and Guhaniyogi, Rajarshi and Scheffler, Aaron and Francom, Devin and Pasqualini, Donatella},
  journal={arXiv preprint arXiv:2503.20528},
  year={2025}
}

@article{wang2007support,
  title={Support vector machine learning-based fMRI data group analysis},
  author={Wang, Ze and Childress, Anna R and Wang, Jiongjiong and Detre, John A},
  journal={NeuroImage},
  volume={36},
  number={4},
  pages={1139--1151},
  year={2007},
  publisher={Elsevier}
}

@article{sung2022functional,
  title={Functional-input gaussian processes with applications to inverse scattering problems},
  author={Sung, Chih-Li and Wang, Wenjia and Cakoni, Fioralba and Harris, Isaac and Hung, Ying},
  journal={arXiv preprint arXiv:2201.01682},
  year={2022}
}

@book{vidakovic2009statistical,
  title={Statistical modeling by wavelets},
  author={Vidakovic, Brani},
  year={2009},
  publisher={John Wiley \& Sons}
}

@article{bliznyuk2008bayesian,
  title={Bayesian calibration and uncertainty analysis for computationally expensive models using optimization and radial basis function approximation},
  author={Bliznyuk, Nikolay and Ruppert, David and Shoemaker, Christine and Regis, Rommel and Wild, Stefan and Mugunthan, Pradeep},
  journal={Journal of Computational and Graphical Statistics},
  volume={17},
  number={2},
  pages={270--294},
  year={2008},
  publisher={Taylor \& Francis}
}

@article{guhaniyogi2024bayesian,
  title={Bayesian data sketching for varying coefficient regression models},
  author={Guhaniyogi, Rajarshi and Baracaldo, Laura and Banerjee, Sudipto},
  year={2024}
}

@article{scheipl2015functional,
  title={Functional additive mixed models},
  author={Scheipl, Fabian and Staicu, Ana-Maria and Greven, Sonja},
  journal={Journal of Computational and Graphical Statistics},
  volume={24},
  number={2},
  pages={477--501},
  year={2015},
  publisher={Taylor \& Francis}
}

@article{kennedy2001calibrate,
  title={Bayesian calibration of computer models},
  author={Kennedy, Marc C and O'Hagan, Anthony},
  journal={Journal of the Royal Statistical Society: Series B (Statistical Methodology)},
  volume={63},
  number={3},
  pages={425--464},
  year={2001},
  publisher={Wiley Online Library}
}

@article{guhaniyogi2022distributed,
  title={Distributed Bayesian varying coefficient modeling using a Gaussian process prior},
  author={Guhaniyogi, Rajarshi and Li, Cheng and Savitsky, Terrance D and Srivastava, Sanvesh},
  journal={The Journal of Machine Learning Research},
  volume={23},
  number={1},
  pages={3642--3700},
  year={2022},
  publisher={JMLRORG}
}

@book{ramsay2005functional,
  title={Functional data analysis},
  author={Ramsay, James O and Silverman, Bernard W},
  year={2005},
  publisher={Springer}
}

@book{ferraty2006nonparametric,
  title={Nonparametric functional data analysis: theory and practice},
  author={Ferraty, Fr{\'e}d{\'e}ric and Vieu, Philippe},
  year={2006},
  publisher={Springer}
}

@article{cao2010future,
  title={The Future of Functional Data Analysis},
  author={Cao, Jiguo and Nielsen, Jason and Ramsay, James and Yao, Fang},
  year={2010}
}

@article{ainsworth2011functional,
  title={Functional data analysis in ecosystem research: the decline of Oweekeno Lake sockeye salmon and Wannock River flow},
  author={Ainsworth, LM and Routledge, R and Cao, J},
  journal={Journal of Agricultural, Biological, and Environmental Statistics},
  volume={16},
  number={2},
  pages={282--300},
  year={2011},
  publisher={Springer}
}

@book{lin2017adaptive,
  title={Adaptive Functional Data Analysis},
  author={Lin, Zhenhua},
  year={2017},
  publisher={University of Toronto (Canada)}
}

@article{guan2020some,
  title={Some new methods and models in functional data analysis},
  author={Guan, Tianyu},
  year={2020},
  publisher={Simon Fraser University}
}

@article{jiang2020filtering,
  title={Filtering-based approaches for functional data classification},
  author={Jiang, Ci-Ren and Chen, Lu-Hung},
  journal={Wiley Interdisciplinary Reviews: Computational Statistics},
  volume={12},
  number={4},
  pages={e1490},
  year={2020},
  publisher={Wiley Online Library}
}

@article{cai2021efficient,
  title={Efficient estimation for varying-coefficient mixed effects models with functional response data},
  author={Cai, Xiong and Xue, Liugen and Pu, Xiaolong and Yan, Xingyu},
  journal={Metrika},
  volume={84},
  number={4},
  pages={467--495},
  year={2021},
  publisher={Springer}
}

@article{sacks1989design,
  title={Design and analysis of computer experiments},
  author={Sacks, Jerome and Welch, William J and Mitchell, Toby J and Wynn, Henry P},
  journal={Statistical science},
  volume={4},
  number={4},
  pages={409--423},
  year={1989},
  publisher={Institute of Mathematical Statistics}
}

@article{bayarri2007computer,
  title={Computer model validation with functional output},
  author={Bayarri, MJ and Walsh, D and Berger, JO and Cafeo, J and Garcia-Donato, G and Liu, F and Palomo, J and Parthasarathy, RJ and Paulo, R and Sacks, J},
  year={2007}
}

@article{conti2010bayesian,
  title={Bayesian emulation of complex multi-output and dynamic computer models},
  author={Conti, Stefano and O’Hagan, Anthony},
  journal={Journal of statistical planning and inference},
  volume={140},
  number={3},
  pages={640--651},
  year={2010},
  publisher={Elsevier}
}

@article{preda2007regression,
  title={Regression models for functional data by reproducing kernel Hilbert spaces methods},
  author={Preda, Cristian},
  journal={Journal of statistical planning and inference},
  volume={137},
  number={3},
  pages={829--840},
  year={2007},
  publisher={Elsevier}
}

@article{kim2018additive,
  title={Additive function-on-function regression},
  author={Kim, Janet S and Staicu, Ana-Maria and Maity, Arnab and Carroll, Raymond J and Ruppert, David},
  journal={Journal of Computational and Graphical Statistics},
  volume={27},
  number={1},
  pages={234--244},
  year={2018},
  publisher={Taylor \& Francis}
}

@article{sun2018optimal,
  title={Optimal penalized function-on-function regression under a reproducing kernel Hilbert space framework},
  author={Sun, Xiaoxiao and Du, Pang and Wang, Xiao and Ma, Ping},
  journal={Journal of the American Statistical Association},
  volume={113},
  number={524},
  pages={1601--1611},
  year={2018},
  publisher={Taylor \& Francis}
}

@article{freund1997decision,
  title={A decision-theoretic generalization of on-line learning and an application to boosting},
  author={Freund, Yoav and Schapire, Robert E},
  journal={Journal of computer and system sciences},
  volume={55},
  number={1},
  pages={119--139},
  year={1997},
  publisher={Elsevier}
}

@article{gelfand2003spatial,
  title={Spatial modeling with spatially varying coefficient processes},
  author={Gelfand, Alan E and Kim, Hyon-Jung and Sirmans, CF and Banerjee, Sudipto},
  journal={Journal of the American Statistical Association},
  volume={98},
  number={462},
  pages={387--396},
  year={2003},
  publisher={Taylor \& Francis}
}

@article{biller2001bayesian,
  title={Bayesian varying-coefficient models using adaptive regression splines},
  author={Biller, Clemens and Fahrmeir, Ludwig},
  journal={Statistical Modelling},
  volume={1},
  number={3},
  pages={195--211},
  year={2001},
  publisher={Sage Publications Sage CA: Thousand Oaks, CA}
}

@article{li2015spatial,
  title={Spatial Bayesian variable selection and grouping for high-dimensional scalar-on-image regression},
  author={Li, Fan and Zhang, Tingting and Wang, Quanli and Gonzalez, Marlen Z and Maresh, Erin L and Coan, James A},
  year={2015}
}

@article{fotheringham1998geographically,
  title={Geographically weighted regression: a natural evolution of the expansion method for spatial data analysis},
  author={Fotheringham, A Stewart and Charlton, Martin E and Brunsdon, Chris},
  journal={Environment and planning A},
  volume={30},
  number={11},
  pages={1905--1927},
  year={1998},
  publisher={Sage Publications Sage UK: London, England}
}

@Article{GWmodel,
    title = {{GWmodel}: An {R} Package for Exploring Spatial
      Heterogeneity Using Geographically Weighted Models},
    author = {{Isabella Gollini} and {Binbin Lu} and {Martin Charlton}
      and {Christopher Brunsdon} and {Paul Harris}},
    journal = {Journal of Statistical Software},
    year = {2015},
    volume = {63},
    number = {17},
    pages = {1--50},
    doi = {10.18637/jss.v063.i17},
}

@Article{spBayes,
    title = {{spBayes}: An {R} Package for Univariate and Multivariate
      Hierarchical Point-Referenced Spatial Models},
    author = {{Andrew O. Finley} and {Sudipto Banerjee} and {Bradley P.
      Carlin}},
    journal = {Journal of Statistical Software},
    year = {2007},
    volume = {19},
    number = {4},
    pages = {1--24},
    url = {https://www.jstatsoft.org/article/view/v019i04},
}

@book{deboor1978splines,
  title={A practical guide to splines},
  author={De Boor, Carl and De Boor, Carl},
  volume={27},
  year={1978},
  publisher={springer New York}
}

\end{document}